\documentclass[conference]{IEEEtran}
\IEEEoverridecommandlockouts

\usepackage{cite}
\usepackage[subtle,tracking=normal]{savetrees} %
\usepackage{amsmath,amssymb,amsfonts}
\usepackage{algorithmic}
\usepackage{graphicx}
\usepackage{textcomp}
\usepackage[table,dvipsnames]{xcolor}
\usepackage[most]{tcolorbox} %
\usepackage{url}
\usepackage{multirow}
\usepackage{booktabs}
\newcommand{\midsepremove}{\aboverulesep=0mm \belowrulesep=0mm}
\midsepremove
\newcommand{\midsepdefault}{\aboverulesep=0mm \belowrulesep=0mm}
\midsepdefault

\usepackage{lipsum}
\usepackage{tcolorbox}
\usepackage[absolute,overlay]{textpos}

\usepackage{tabularx}
\usepackage{balance}
\usepackage{bm}

\def\BibTeX{{\rm B\kern-.05em{\sc i\kern-.025em b}\kern-.08em
    T\kern-.1667em\lower.7ex\hbox{E}\kern-.125emX}}

\begin{document}

\title{Augur: Pre-Execution Energy Prediction for\\ Workflow Tasks in Heterogeneous Clusters\\
}

\author{
    \IEEEauthorblockN{
        Kathleen West\textsuperscript{a}, Vasilis Bountris\textsuperscript{b}, Philipp Thamm\textsuperscript{b}, Ulf Leser\textsuperscript{b}, Yehia Elkhatib\textsuperscript{a}, Lauritz Thamsen\textsuperscript{a}
    }
    \IEEEauthorblockA{
        \textsuperscript{a}University of Glasgow, United Kingdom \\
        \textsuperscript{b}Humboldt-Universität zu Berlin, Germany \\
    }
}
\maketitle

\begin{textblock*}{\textwidth}(17.0mm,260mm)
    \begin{tcolorbox}[width=\textwidth, colback=gray!10, colframe=gray!10, sharp corners, boxrule=0.5pt, boxsep=2pt]
        \centering \small \textbf{For the purpose of open access, we have applied a Creative Commons Attribution (CC BY) license to this version of our paper.}
    \end{tcolorbox}
\end{textblock*}

\begin{abstract} 
Scientific workflows are widely used to process large quantities of data, leading to significant energy consumption and carbon emissions. To reduce this environmental impact, energy and carbon-aware scheduling approaches could be employed. However, such methods require runtime and energy predictions, which are typically only available for workflows that have been executed previously. 
Meanwhile, scientists may execute new or modified workflows, use workflows with different input data, or run them on alternative infrastructure. 
To address this critical gap, we propose Augur, a novel method to predict the energy consumption of scientific workflow tasks prior to execution. By efficiently profiling both the available cluster infrastructure and the workflow at hand, Augur is capable of predicting the overall energy consumption of the workflow with a median prediction error of $\bm{16.3\% \pm 15.3\%}$ compared to Ichnos, an energy estimation method that uses fitted power models, and $\bm{18.2 \pm 14.7\%}$ compared to Intel RAPL, as observed in our experimental evaluation on public and private cloud infrastructure. Relying on only minimal historical execution data, Augur outperforms two state-of-the-art methods in predicting both task runtime and total workflow energy, providing a robust foundation for energy-efficient and carbon-aware scientific data analysis.
\end{abstract}


\section{Introduction}

Scientists in many fields, including genomics, materials science, and remote sensing, need to analyze increasing amounts of data \cite{muirRealCostSequencing2016b, fellowsyatesReproduciblePortableEfficient2021, schaarschmidtWorkflowEngineeringMaterials2021, berrimanMontageGridEnabled2004}. 
Scientific workflow systems, like Nextflow~\cite{ditommasoNextflowEnablesReproducible2017a}, facilitate the automation of such analyses, enabling scientists to create pipelines from tasks and data dependencies. 
Because these workflows are often used to process large quantities of data, they tend to be resource-intensive and long-running, leading to significant energy consumption and, therefore, carbon emissions. 
At scale, workflows contribute to a broader trend, where the growing prevalence of big data applications has generally been identified as a driver of the increasing emissions of the ICT sector~\cite{freitagRealClimateTransformative2021b}. As such, it is crucial to quantify and reduce the energy consumption and carbon emissions of workloads such as scientific workflows in order to reduce their impact~\cite{vivas2024trends, da2024workflows}. 

When scientific workflows are executed on heterogeneous infrastructure, they can be scheduled using methods that minimize a workflow's energy consumption or carbon emissions~\cite{reddy2023multi, durillo2014multi, bostandoost2025quantifyingcarbonreductiondag, west2026systematic, 10.1145/3754598.3754673, 9298863}. 
However, the existing energy or carbon-aware workflow scheduling methods simply assume that they have prior knowledge of workflow structures, task runtimes, or energy consumption~\cite{bostandoost2025quantifyingcarbonreductiondag, west2026systematic, 10.1145/3754598.3754673, 9298863}, or they make use of ML-based prediction methods that rely on having large amounts of historical workflow execution data to effectively schedule workflows~\cite{durillo2015pareto, durillo2014multi}.

An alternative approach is to predict workflow task characteristics through efficient profiling.
Such methods have shown promise in predicting task runtimes and memory requirements from profiling workflows on input samples and on limited resources~\cite{bader2024lotaru, da2015online, rosa2018cost, rosa2021computational}, yet this approach has not been explored for predicting energy consumption or carbon emissions.
Instead, prior work on simulating~\cite{da2020characterizing} and predicting~\cite{queirozperformance} workflow energy consumption assumes the availability of a wealth of training data for ML-based prediction models, like the workflow scheduling approaches. 
These methods rely on the assumption that new workflow executions exhibit behavior similar to previous runs, which often does not hold in practice: execution characteristics can vary substantially given different input data, dynamic workflow plans, and heterogeneous compute resources.
In particular, the processor and memory hardware significantly affects runtimes, energy consumption, and carbon emissions~\cite{west2026systematic}.
Moreover, some methods only estimate energy consumption at the workflow-level~\cite{queirozperformance}. 

We argue that workflow energy prediction approaches must provide efficient task-level estimates while relying on only limited prior execution data. 
We therefore propose, Augur, an approach for predicting the energy consumption of scientific workflow tasks on heterogeneous clusters using single-node profiling. A single node is used to execute workflow micro-benchmarks to estimate task runtime and CPU utilization under full-size inputs.
This profiling node does not need to be part of the target cluster and can instead be an external resource, such as a standalone or a cloud server.
Augur then combines the single-node predictions with infrastructure profiling data from target cluster nodes to estimate the energy consumption of workflow tasks across heterogeneous resources.
To account for dynamic workflows, where execution paths may vary with input data, Augur incorporates a reflector module that provides fallback predictions for previously unseen tasks. 

We contribute the following: 
\begin{itemize}
    \item the design of Augur, a single-node profiling approach for task-level workflow energy prediction (Section~\ref{sec:system-design});
    \item an evaluation using four Nextflow core-library workflows and four Intel servers (Section~\ref{sec:evaluation}); and
    \item an open-source implementation of Augur: 
\end{itemize}

\url{https://github.com/GlasgowC3lab/augur}

\section{Background} 
We explain scientific workflows, and the estimation of energy consumption and carbon emissions. 

\subsection{Scientific Workflows} 

\begin{figure}[h]
\centerline{\includegraphics[width=\columnwidth]{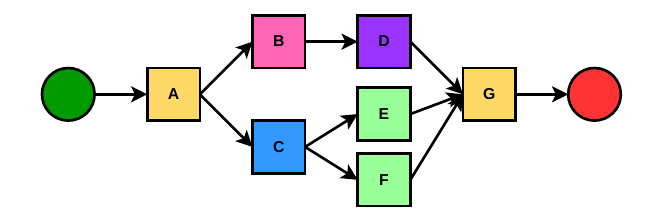}}
\caption{A scientific workflow formed of seven tasks.}
\label{fig:simple-workflow}
\end{figure}

Scientific workflows are typically depicted as directed acyclic graphs (DAGs). In these graphs, nodes represent computational tasks, and edges illustrate the data or control dependencies between them.
Scientific workflow systems automate the execution of workflows. 
Fig.~\ref{fig:simple-workflow} shows an example workflow, consisting of seven tasks that depend on each other; e.g., Tasks E and F require input from Task C.

These tasks are typically self-contained programs that are often shared as containers. Individual tasks are considered atomic and can be executed on different machines with different CPUs and attached memory, leading to changes in runtime and energy consumption. The same task can be executed with different input data, which also changes the runtime and energy consumption. 

In this work, we define \textit{heterogeneous} clusters as comprising nodes equipped with varied configurations; e.g. differing CPU architectures, CPU frequencies, and memory capacities. These variations lead to different energy consumption profiles.  

\subsection{Workflow Task Energy Consumption}
The overall energy consumption of a scientific workflow is defined as the sum of energy consumed by all of its constituent tasks, which may be distributed across different machines. 
Since the CPU and memory subsystems account for the primary load-dependent portion of system energy consumption, this analysis focuses on their consumption. In contrast, the energy consumption of local network infrastructure is substantially less dependent on load and is excluded from consideration. 

\begin{figure}[h]
\centerline{\includegraphics[width=0.6\columnwidth,trim={20 10 20 10},clip]{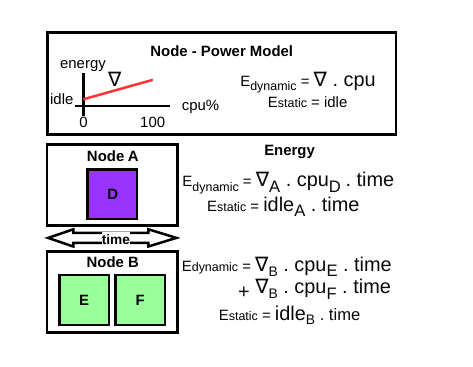}}
\caption{How energy consumption can be estimated on multiple nodes.}
\label{fig:workflow-energy}
\end{figure}

Figure~\ref{fig:workflow-energy} shows that for each node, we calculate the energy consumed by tasks running on CPUs by summing their static and dynamic energy, using a linear power model. 
The static energy is the energy consumed when the node is at 0\% CPU utilization -- the power model's intercept. This is multiplied by the length of time when any number of tasks are running, on each node. 
The dynamic energy is estimated using the gradient ($\nabla$) of the power model multiplied by the task CPU utilization and runtime. This is calculated for each individual task.  

The memory energy consumption is found by multiplying the fraction of memory allocated to a specific task with the memory coefficient and runtime. The memory coefficient (W/GB) is constant and measured for each individual machine, where possible.

\section{System Design} %
\label{sec:system-design}
Augur's purpose is to predict workflow task energy consumption across all nodes in a heterogeneous cluster, prior to a workflow's execution. It achieves that by profiling the cluster infrastructure and profiling the workflow on a single node with small samples of the input data. The resulting profiling task runtime and resource usage metrics are used to predict workflow task runtimes, CPU utilization, and energy consumption for the full-sized execution. Moreover, a reflector module is used to predict energy consumption for tasks that remain unseen in profiling runs, but are present in the full workflow using minimal historical execution data.

These predictions could then be used by a resource manager to schedule a workflow on a heterogeneous cluster, e.g. to minimize energy consumption or carbon emissions.

\subsection{Approach Overview}
\begin{figure}[htbp]
\centerline{\includegraphics[width=\columnwidth,trim={20 20 110 10},clip]{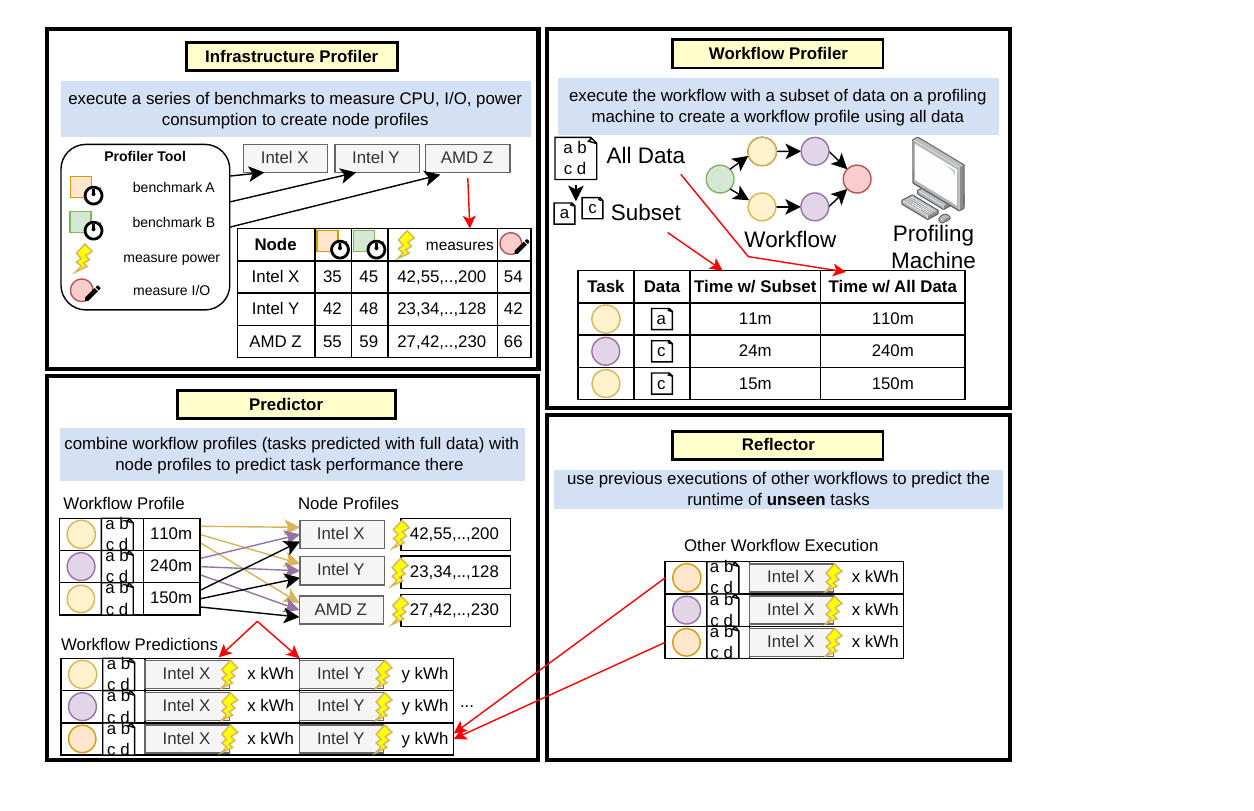}}
\caption{High-level system architecture of Augur, comprising the infrastructure profiler, workflow profiler, predictor, and reflector modules.}
\label{fig:augur-design}
\end{figure}

Augur consists of four components, depicted in Fig.~\ref{fig:augur-design}, which are explained in the following sections. 

\begin{itemize}
\item The infrastructure profiler (Section~\ref{sec:infra-profiler}) collects performance metrics for heterogeneous infrastructure where the workflows are executed and the profiling node. 
\item The workflow profiler (Section~\ref{sec:workflow-profiler}) is used to execute the targeted workflow on the profiling node with a subset of input data to gain insights into the behavior of workflow tasks. 
\item The predictor (Section~\ref{sec:predictor}) combines the infrastructure profiling metrics with the collected insights into workflow behavior to predict the runtime and energy consumption of workflow tasks on targeted cluster node. 
\item The reflector (Section~\ref{sec:reflector}) uses previous workflow executions as a fallback for workflow tasks unseen during profiling. 
\end{itemize}

\subsection{Assumptions}
\label{sec:assumptions}
We make the following four assumptions:
\begin{enumerate}
    \item Scientific workflows can be executed using down-sampled data from the full sized input dataset.
    \item Many scientific workflow task runtimes scale linearly as input data size increases.
    \item Scientists have access to a profiling machine, with sufficient memory to run workflow profiling.
    \item Scientists have the ability to perform infrastructure profiling regularly, or when hardware changes.
\end{enumerate}

\subsection{Infrastructure Profiling}
\label{sec:infra-profiler}
The infrastructure profiling module runs on the single node used for workflow profiling, and at least each type of node of a heterogeneous target cluster to gather performance metrics. 
It first gathers general information around the type of machine utilized, i.e. the number of cores, the amount of RAM. It then executes a series of benchmarks to measure CPU performance with sysbench\footnote{\url{https://github.com/akopytov/sysbench}\label{mark:sysbench}} and z7b\footnote{\url{https://www.7-zip.org/}\label{mark:z7b}}, and I/O with fio\footnote{\url{https://github.com/axboe/fio}\label{mark:fio}}. It furthermore executes a Nextflow benchmark workflow, generated using WfBench\footnote{\url{https://github.com/wfcommons/WfCommons} \label{mark:wfbench}} to gain insights into Nextflow performance on utilized infrastructure. 

For the nodes where the workflow is intended to run, power measurements are taken to generate a power model, used to predict energy consumption. This is not required for the node used to profile workflows, and if direct measurements are not available, e.g. on a public cloud instance, an approximate model could be created from estimated power draws\footnote{\url{https://www.cloudcarbonfootprint.org/docs/methodology/}\label{mark:ccf}}.

The infrastructure profiling module takes around ten minutes to run, and it can be run in parallel over heterogeneous target cluster nodes. 
The profiling should be performed regularly, when cluster resource management systems detect a change to compute resources, or at regular intervals to track performance degradation over time. 

\subsection{Workflow Profiling}
\label{sec:workflow-profiler}
We use a single node to perform workflow profiling, which must have sufficient memory to run the selected scientific workflow with down-sampled data. 
The down-sampled data should be substantially smaller than the full-sized inputs, e.g. 5--10\% of the full dataset, and should be created so that it allows full workflow execution, possibly using available splitter tools such as seqkit\footnote{\url{https://github.com/shenwei356/seqkit}}. The workflow would be executed on the profiling machine with at least three samples, each of a different size.
This allows Augur to check whether task runtimes are correlated with the input size, when generating task prediction models. 

Once executed, the automatically generated workflow traces would be collected, along with a log of the input data used by each workflow task (generated by enabling the nf-datatrail plugin\footnote{\url{https://github.com/Lehmann-Fabian/nf-datatrail}\label{mark:datatrail}}). 

\subsection{Prediction of Task Energy Consumption}
\label{sec:predictor}
The prediction module combines insights gained from infrastructure profiling and workflow profiling to predict the workflow task runtime and, subsequently, energy consumption for targeted infrastructure.

\subsubsection{Predicting Task Runtimes}
First, the workflow profiling runs are processed, where each unique task is identified, and the profiling data collated. For each task, we calculate the Pearson correlation coefficient between the task input data size and the task runtime. We choose between three different prediction models -- used to predict the runtime of the task, given the size of input data.

\begin{itemize}
    \item Case 1: When we do not find a significant relationship between the input data size and task runtime (Pearson value is less than 0.5), we predict the task runtime as the median runtime of the task.
    \item Case 2: When we identify that there is a weak linear relationship between the input data size and task runtime, we opt to use a Gradient Boosting regression model which can capture non-linearity in a relationship and performed best in our experiments.
    \item Case 3: When we identify a strong linear relationship between task input size and runtime, we use a Linear Regression model. 
\end{itemize}

With these models, we predict the runtime of the workflow task with the full size input data on the profiling machine.

\subsubsection{Translating Task Runtime from Local to Target Node}
Before making any energy consumption predictions, the full-sized task runtime predictions must be adjusted, from the profiling machine which we refer to as the \textit{local node}, for the targeted cluster node. 

For the runtime, we generate a translation factor by comparing the profiling node's infrastructure metrics with the target machine's infrastructure profile.

We calculate the CPU ratio between the profiling and target machine using the duration, $t$, of a CPU benchmark program. If the profiling node is more powerful than the target machine, i.e. it has more cores, we scale the CPU ratio using the number of cores, $c$. 
\begin{equation}
r_{cpu} = 
    \begin{array}{ll}
        \frac{(\frac{t_{local}}{t_{target}} + \frac{c_{local}}{c_{target}})}{2} & \mbox{if } c_{local} > c_{target} \\
        \frac{t_{local}}{t_{target}} & \mbox{otherwise}
    \end{array}
\end{equation}

For the memory ratio, we calculate the ratio between the profiling and target machine's I/O write rate, $io\_w$. 
\begin{equation}
r_{mem} = \frac{io\_w_{local}}{io\_w_{target}}
\end{equation}

Furthermore, if the local machine has a different number of cores compared to the target machine, we weigh the CPU ratio more heavily in our translation factor, $f$. In our experience, this has a greater impact than nodes having access to more/less memory. Otherwise, we evenly weigh the CPU and memory in our factor. 
\begin{equation}
f =
	\begin{array}{ll}
		0.7 * r_{cpu} + 0.3 * r_{mem}  & \mbox{if } c_{local} \ne c_{target} \\
		0.5 * r_{cpu} + 0.5 * r_{mem} & \mbox{otherwise}
	\end{array}
\end{equation}

\subsubsection{Predicting Task Energy Consumption}
To predict energy consumption, we also need to predict CPU utilization on target compute nodes. Each node that we use has a power model that is generated from infrastructure profiling in the form:
\begin{equation}
    E = \nabla_{node} (CPU_{usage}) + idle_{node}
\end{equation}

We have observed that the CPU utilization of a task during profiling is often similar to the utilization using the full dataset on the same infrastructure. Therefore, we expect the CPU utilization of the full-sized task to be the same as that of the task running with down-sampled data ($CPU_{usage}$). The predicted task runtime is referred to as $t_{pred}$. 
Then, each task's dynamic CPU energy consumption can be predicted with:
\begin{equation}
E_{dynamic} = t_{pred} * \nabla_{node} (CPU_{usage})
\end{equation}

The static CPU energy consumption can be calculated over the duration that one or more tasks execute on each node %
with:
\begin{equation}
E_{static} = duration * idle_{node}
\end{equation}

We calculate the memory energy consumption, for each task, using the predicted time taken to run a task, $t_{pred}$, the size of memory allocated to the task, $mem_{size}$, and the memory energy coefficient, $mem_{coeff}$, which is constant and measured per node. 
\begin{equation}
E_{memory} = t_{pred} * mem_{size} * mem_{coeff}
\end{equation}

Augur makes predictions of each task's dynamic energy consumption, and the overall workflow energy consumption once static energy has been calculated.

\subsection{Reflection from Previous Executions}
\label{sec:reflector}
Profiling a workflow with down-sampled data typically results in incomplete coverage of tasks, %
because workflow execution plans are often created dynamically based on input characteristics. This could be due to improper down-sampled data generation, i.e. missing out on some part of input, but it might not always be possible to generate down-sampled input data that represents the original data and allows to fully run the workflow with all possible tasks being executed.

To reduce the likelihood of encountering new tasks without a prediction model, we developed the reflector module. This module assumes that scientists use Augur to make predictions for a (shared) cluster where scientific workflows have been executed. Traces for these executions are available, and serve as historical data that can be used to train models for tasks -- given that workflow tasks can be reused across workflows. This ensures that we can make predictions for tasks that remain ``unseen'' during profiling.

\section{Evaluation} 
\label{sec:evaluation}
We now describe the experimental setup (Section~\ref{subsec:setup}) that we used in order to assess:%
\begin{itemize}
  \item how accurately the workflow and task energy consumption can be predicted, compared to Ichnos, an energy estimation method using fitted power models (Section~\ref{subsec:predict-energy-consumption}),
  \item the impact of having limited historical data from other workflows available to Augur (Section~\ref{subsec:reflector-analysis}),
  \item how workflow energy consumption estimates compare to RAPL energy readings (Section~\ref{subsec:augur-vs-rapl}),
  \item how well our workflow task runtime predictions compare against Lotaru~\cite{bader2024lotaru}, a state-of-the-art workflow task runtime prediction method (Section~\ref{subsec:predict-task-runtime}),
  \item how our workflow energy estimates compare against Barbosa et al's~\cite{queirozperformance}, a state-of-the-art method for predicting workflow energy consumption (Section~\ref{subsec:vs-baseline}),
  \item the overhead of our approach (Section~\ref{subsec:profiling-overhead}),
  \item and the challenges involved (Section~\ref{subsec:discussion}).
\end{itemize}

\subsection{Experimental Setup}
\label{subsec:setup}

\paragraph{Infrastructure}
We use compute nodes of four different architectures, as detailed in Table~\ref{table:compute-resources}. The environment covers both shared compute clusters and cloud instances. 
The local infrastructures consist of two homogeneous commodity clusters, formed from 4 Neptune and 4 Jupiter nodes, respectively. The setup is augmented with two cloud-based nodes: an AWS metal \texttt{c5d} instance (designated as Diana), and a general-purpose AWS \texttt{m5.4xlarge} instance (Minerva). 

\begin{table}[htb]
\caption{The compute resources used in the study.}
\label{table:compute-resources}
\centering
\setlength{\tabcolsep}{5pt}%
\begin{tabular}{llccl}
\toprule
& & Cores & Memory & \\
Node & Hardware & (\#) & (GB) & Type \\
\toprule
Neptune & Intel Xeon E5-2640 & 32 & 64 & Cluster \\ %
Jupiter & Intel Xeon Silver 4314 & 32 & 256 & Cluster \\ %
Diana & Intel Xeon Scalable Processor & 96 & 192 & Cloud \\
Minerva & Intel Xeon Scalable Processor & 16 & 64 & Cloud \\
\bottomrule
\end{tabular}
\end{table}

\paragraph{Infrastructure Profiling}
When undertaking infrastructure profiling, we document the number of cores, available RAM, using fio\textsuperscript{\ref{mark:fio}} to measure I/O read and write performance, and the time taken to run the sysbench\textsuperscript{\ref{mark:sysbench}} and z7b\textsuperscript{\ref{mark:z7b}} benchmarks to measure CPU performance. 
We generate a benchmark workflow\textsuperscript{\ref{mark:wfbench}}, and execute it with Nextflow to gain insights into system overhead. We also take energy measurements, used to generate a power model for each utilized node. The infrastructure profiling takes around ten minutes to run and can be parallelized across nodes. It should be repeated when a resource manager detects any change to infrastructure, or at regular intervals to track performance degradation over time. This can be performed by system administrators, or those with privileged access. 

\begin{table}[htb]
\caption{Excerpt of infrastructure profiling results.}
\label{table:infrastructure-profiling}
\centering
\begin{tabular}{lccccc}
\toprule
& RAM & CPUs & IO\_{read} & IO\_{write} & z7b \\
Node & (GB) & (\#) & (MiB/s) & (MiB/s) & (s) \\
\midrule
Neptune & 64 & 32 & 109 & 91.7 & 1779 \\
Jupiter & 256 & 32 & 1362 & 952 & 2711 \\ 
Diana & 192 & 96 & 127 & 127 & 2973 \\
Minerva & 64 & 16 & 127 & 127 & 2963  \\
\bottomrule
\end{tabular}
\end{table}

\paragraph{Workflows}
We selected workflows from nf-core, a community-curated collection of scientific workflows created using Nextflow~\cite{ewels2020nf}. Specifically, we executed four of the ten most used bioinformatics workflows:
\begin{itemize}
    \item RNA-Seq -- an analysis pipeline for RNA sequencing data obtained from organisms
    \item Chip-Seq -- an analysis pipeline for Chromatin ImmunoPrecipitation sequencing data
    \item Nano-Seq -- an analysis pipeline for Nanopore DNA/RNA sequencing data
    \item Atac-Seq -- an analysis pipeline for ATAC-seq data
\end{itemize}
Bioinformatics was selected as the primary application domain as it is representative of the majority of nf-core workflows.

\paragraph{Workflow Profiling}
We down-sampled the full-size input data for each of the studied workflows, where the down-sampled data would be up to 4\% of the full input dataset. We collected at least three samples, which we used to conduct workflow profiling runs where we collect the workflow trace data, recording each task's runtime and resource utilization. 
We used the nf-datatrail plugin\textsuperscript{\ref{mark:datatrail}} to collect additional information around individual task file input and output data. 
We discuss the overhead of workflow profiling in Section~\ref{subsec:profiling-overhead}. 

\paragraph{Ground Truth Workflow Energy Consumption}
We executed each workflow three times on the nodes of the Neptune and Jupiter clusters, for which we report the mean metrics.

Ichnos~\cite{west2025ichnoscarbonfootprintestimator} is our state-of-the-art tool used to estimate the energy consumption and carbon emissions of Nextflow workflow executions from traces. It uses fitted linear power models to estimate the energy consumption, resulting in 
estimates within $5.3\pm6.3$\% of RAPL measurements, polling at 1ms intervals. 
These RAPL measurements encompass all energy consumed by CPU and memory~\cite{khan2018rapl}, which may include background services other than the workflow itself.
However, RAPL readings taken for a node cannot distinguish between concurrent task executions to directly estimate workflow energy consumption at the task-level.
This would require approximating task energy using utilisation metrics~\cite{thamm2025strategiesmeasureenergyconsumption}, which would no longer be a measured ground truth. 

We, therefore, first compare Augur task-level energy predictions with Ichnos' task-level dynamic energy estimations and workflow-level energy consumption, and later compare Augur's workflow-level energy prediction with RAPL readings. 

\paragraph{Comparison with Baseline Methodologies}
To effectively evaluate how well we predict workflow task runtime and energy consumption, we compare our approach also with state-of-the-art baselines, in addition to comparing our predictions with the ground truth estimations and measurements in Sections~\ref{subsec:predict-energy-consumption}~and~\ref{subsec:augur-vs-rapl}.
We begin by comparing our workflow task runtime predictions with Lotaru, a state-of-the-art profiling-based method to predict workflow task runtime, in Section~\ref{subsec:predict-task-runtime}.
Afterwards, we compare our workflow energy consumption predictions with Barbosa's work, which identified an approach to predict energy consumption at a workflow-level using workflow execution parameters, in Section~\ref{subsec:vs-baseline}.

\subsection{Prediction of Energy Consumption}
\label{subsec:predict-energy-consumption}
We show Augur's predicted energy consumption in comparison to Ichnos in Table~\ref{table:augur-energy-vs-ichnos}. These predictions are made with the reflector module enabled. 

Here, we first focus on the workflow \textit{dynamic energy}, which is the sum of each individual task's dynamic CPU and memory energy. We compare the predictions of Augur with that of Ichnos, and report the percentage error between them as the \textit{dynamic prediction error}. We also took the percentage error between each task's predicted dynamic energy and Ichnos', and report the mean as the \textit{task energy prediction error}. Finally, we consider the workflow's \textit{total energy}, which is the sum of all workflow task's dynamic CPU and memory energy, and each node's static CPU energy. We compare Augur's prediction with Ichnos', and report the percentage error between them as the \textit{total prediction error}.

\begin{table*}[htb]
\caption{Using Augur to predict workflow dynamic and overall energy consumption, compared against Ichnos.} 
\label{table:augur-energy-vs-ichnos}
\centering
\begin{tabular}{ll|ccc||c||ccc}
\toprule
Profiling& & \multicolumn{3}{c||}{Workflow Dynamic Energy} & Task-Level & \multicolumn{3}{c}{Workflow Total Energy} \\
Node to & & Augur & Ichnos & Prediction & Dynamic Energy & Augur & Ichnos & Prediction \\
Target & Workflow & (kWh) & (kWh) & Error (\%) & Prediction Error (\%) & (kWh) & (kWh) & Error (\%) \\
\midrule
& RNA-Seq & 0.40 & 0.33 & +21.70 & 56.87 & 1.35 & 1.21 & \cellcolor{Green!30}+11.56 \\
Jupiter to& Chip-Seq & 0.42 & 0.47 & -11.00 & 101.60 & 1.38 & 1.61 & \cellcolor{Green!30}-14.59 \\
Neptune& Nano-Seq & 0.12 & 0.20 & -39.59 & -0.03 & 0.41 & 0.70 & \cellcolor{Red!30}-41.16 \\
& Atac-Seq & 0.39 & 0.33 & +21.09 & 48.66 & 1.27 & 1.13 & \cellcolor{Green!30}+12.82 \\
\midrule
& RNA-Seq & 1.41 & 1.54 & -8.23 & 14.10 & 2.53 & 2.65 & \cellcolor{Green!30}-4.49 \\
Neptune to& Chip-Seq & 2.87 & 3.13 & -8.59 & -2.40 & 4.99 & 5.43 & \cellcolor{Green!30}-8.09 \\
Jupiter& Nano-Seq & 0.06 & 0.10 & -38.01 & 9.23 & 0.19 & 0.39 & \cellcolor{Red!30}-52.72 \\
& Atac-Seq & 0.17 & 0.17 & +1.14 & 44.66 & 0.43 & 0.59 & \cellcolor{Yellow!30}-27.12 \\
\midrule
& RNA-Seq & 0.28 & 0.33 & -15.14 & 23.99 & 1.02 & 1.21 & \cellcolor{Green!30}-15.96 \\
Diana to& Chip-Seq & 0.34 & 0.47 & -27.88 & 36.38 & 1.17 & 1.61 & \cellcolor{Yellow!30}-27.29 \\
Neptune& Nano-Seq & 0.11 & 0.20 & -45.40 & -7.80 & 0.37 & 0.70 & \cellcolor{Red!30}-46.78 \\
& Atac-Seq & 0.23 & 0.33 & -28.34 & -4.04 & 0.78 & 1.13 & \cellcolor{Red!30}-30.92 \\
\midrule
& RNA-Seq & 1.54 & 1.54 & +0.29 & 62.19 & 2.66 & 2.65 & \cellcolor{Green!30}+0.31 \\
Diana to& Chip-Seq & 3.38 & 3.13 & +7.69 & 61.21 & 5.51 & 5.43 & \cellcolor{Green!30}+1.48 \\
Jupiter& Nano-Seq & 0.10 & 0.10 & +4.86 & 51.29 & 0.28 & 0.39 & \cellcolor{Yellow!30}-29.09 \\
& Atac-Seq & 0.21 & 0.17 & +24.14 & 74.17 & 0.52 & 0.59 & \cellcolor{Green!30}-10.61 \\
\midrule
& RNA-Seq & 0.25 & 0.33 & -21.94 & 20.09 & 0.96 & 1.21 & \cellcolor{Yellow!30}-20.88 \\
Minerva to& Chip-Seq & 0.47 & 0.47 & -1.41 & 58.95 & 1.51 & 1.61 & \cellcolor{Green!30}-5.96 \\
Neptune& Nano-Seq & 0.11 & 0.20 & -44.84 & 23.19 & 0.38 & 0.70 & \cellcolor{Red!30}-45.29 \\
& Atac-Seq & 0.29 & 0.33 & -9.69 & 11.59 & 0.96 & 1.13 & \cellcolor{Green!30}-14.71 \\
\midrule
& RNA-Seq & 1.04 & 1.54 & -32.42 & 10.40 & 2.09 & 2.65 & \cellcolor{Yellow!30}-21.03 \\
Minerva to& Chip-Seq & 3.29 & 3.13 & +4.98 & 16.40 & 5.95 & 5.43 & \cellcolor{Green!30}+9.62 \\
Jupiter& Nano-Seq & 0.07 & 0.10 & -26.05 & 27.83 & 0.21 & 0.39 & \cellcolor{Red!30}-47.58 \\
& Atac-Seq & 0.19 & 0.17 & +14.24 & 45.34 & 0.49 & 0.59 & \cellcolor{Yellow!30}-16.69 \\
\bottomrule
\end{tabular}
\end{table*}

We see that Augur can predict the overall workflow energy consumption with a median prediction error of $16.3\% \pm 15.3\%$ (p75 of 29.5\% and p95 of 47.5\%). 

We observed that the Nano-Seq workflow was consistently predicted with the greatest error, consistently over $29\%$. Upon investigating this, we found that the profiling runs were missing 3 tasks that appeared in the full sized workflow run. In the full run, one such task was expected to run for one hour and was responsible for a significant fraction of overall energy consumption. We believe that instrumenting a down-sampled dataset that captures this task would mitigate this error and allow for greater accuracy. Furthermore, the Nano-Seq workflow is largely formed of tasks that last under five minutes. Consequently, the missing task has a significant impact on the overall prediction. 
Without Nano-Seq, our predictions would have a median error of 13.7\% (p75 of 19.8\% and p95 of 27.8\%) when compared to Ichnos.

These energy consumption predictions were made using node-specific fitted CPU power models obtained from the infrastructure profiling phase, alongside the constant memory coefficient of $0.392 W/GB$. Augur and Ichnos can alternatively be configured with measured node-specific memory coefficients, which slightly alters the energy prediction error, to $17.8 \pm 15.3\%$ when compared to Ichnos.

We observed that the \textit{task energy prediction error} was often higher than the workflow-level prediction error. This is because task-level predictions focus on the dynamic portion of CPU energy consumption; this does not consider the static energy consumed by utilized nodes, which often dominates overall workflow energy consumption.

\subsection{Impact of the Reflector Module}
\label{subsec:reflector-analysis}
We explore the impact of enabling our reflector module, which deals with the case where we encounter an unseen workflow task. That is, a task generated in the full workflow run that was not present in the profiling runs. In this case, we look towards previous workflow executions made on the targeted infrastructure to find the unseen task, and, if found, use the data to make our prediction. These are previous executions of different workflows, which are limited. 
In our experiment, we had 3 executions each of 3 different workflows. 

\begin{table}[htb]
\caption{Comparison of predicted total workflow energy consumption using Augur (with and without the Reflector module) against Ichnos ground truth data.}
\label{table:augur-reflector-analysis}
\centering
\begin{tabular}{lc|ccc}
\toprule
Profiling& & \multicolumn{3}{c}{Energy Prediction Error (\%)} \\
Node to & & \multicolumn{2}{c}{Reflector} & \\
Target & Workflow & Disabled & Enabled & Change \\
\midrule
& RNA-Seq & 38.85 & 11.56 & \cellcolor{Green!30}-27.29 \\
Jupiter to&Chip-Seq & 7.76 & -14.59 & \cellcolor{Red!30}+6.83 \\
Neptune& Nano-Seq & -19.21 & -41.16 & \cellcolor{Red!30}+21.95 \\
& Atac-Seq & 48.53 & 12.82 & \cellcolor{Green!30}-35.71 \\
\midrule
& RNA-Seq & -26.18 & -4.49 & \cellcolor{Green!30}-21.69 \\
Neptune to&Chip-Seq & -29.79 & -8.09 & \cellcolor{Green!30}-21.7 \\
Jupiter& Nano-Seq & -67.26 & -52.72 & \cellcolor{Green!30}-14.54 \\
& Atac-Seq & -37.16 & -27.12 & \cellcolor{Green!30}-10.04 \\
\midrule
& RNA-Seq & -46.12 & -15.96 & \cellcolor{Green!30}-30.16 \\
Diana to&Chip-Seq & -57.91 & -27.29 & \cellcolor{Green!30}-30.62 \\
Neptune& Nano-Seq & -67.01 & -46.78 & \cellcolor{Green!30}-20.23 \\
& Atac-Seq & -48.19 & -30.92 & \cellcolor{Green!30}-17.27 \\
\midrule
& RNA-Seq & -61.55 & 0.31 & \cellcolor{Green!30}-61.24 \\
Diana to&Chip-Seq & -63.87 & 1.48 & \cellcolor{Green!30}-62.39 \\
Jupiter& Nano-Seq & -79.81 & -29.09 & \cellcolor{Green!30}-50.72 \\
& Atac-Seq & -69.32 & -10.61 & \cellcolor{Green!30}-58.71 \\
\midrule
& RNA-Seq & -28.52 & -20.88 & \cellcolor{Green!30}-7.64 \\
Minerva to&Chip-Seq & -21.38 & -5.96 & \cellcolor{Green!30}-15.42 \\
Neptune& Nano-Seq & -52.01 & -45.29 & \cellcolor{Green!30}-6.72 \\
& Atac-Seq & -22.43 & -14.71 & \cellcolor{Green!30}-7.72 \\
\midrule
& RNA-Seq & -48.77 & -21.03 & \cellcolor{Green!30}-27.74 \\
Minerva to&Chip-Seq & -32.65 & 9.62 & \cellcolor{Green!30}-23.03 \\
Jupiter& Nano-Seq & -70.73 & -47.58 & \cellcolor{Green!30}-23.15 \\
& Atac-Seq & -52.83 & -16.69 & \cellcolor{Green!30}-36.14 \\
\bottomrule
\end{tabular}
\end{table}

When the reflector module is enabled, we see a consistent improvement in energy prediction accuracy. In Table~\ref{table:augur-reflector-analysis}, we compare the workflow-level energy prediction error when the reflector module is disabled to when it is enabled, and show the change in overall accuracy. The reflector module can improve accuracy by as much as 62.4\%, and a median improvement of 22.4\%.

\subsection{Comparing Augur with RAPL Readings}
\label{subsec:augur-vs-rapl}
We now compare Augur's predicted energy consumption for the entire workflow with RAPL readings, which provides accurate energy consumption readings for CPU and memory~\cite{khan2018rapl}. Note, however, that RAPL counters track all energy consumed by compute nodes, which can include background processes not related to the workflow. In our experiments, we reserved all compute nodes utilized and did not run any other tasks to minimize background energy consumption. Augur only considers energy consumption attributed to workflow task consumption. Therefore, we expect to generally under-predict workflow energy consumption. 
We show the results in Table~\ref{table:augur-energy-vs-rapl}. 

\begin{table}[htb]
\caption{Comparison of predicted total workflow energy consumption using Augur against hardware-level measurements (RAPL).}
\label{table:augur-energy-vs-rapl}
\centering
\begin{tabular}{ll|ccc}
\toprule
Profiling& & \multicolumn{3}{c}{Workflow Energy} \\
Node to & & Augur & RAPL & Prediction \\
Target & Workflow & (kWh) & (kWh) & Error (\%) \\
\midrule
& RNA-Seq & 1.35 & 1.15 & \cellcolor{Green!30}+17.78 \\
Jupiter to& Chip-Seq & 1.38 & 1.54 & \cellcolor{Green!30}-10.57 \\
Neptune& Nano-Seq & 0.41 & 0.69 & \cellcolor{Red!30}-40.47 \\
& Atac-Seq & 1.27 & 1.08 & \cellcolor{Green!30}+18.00 \\
\midrule
& RNA-Seq & 2.53 & 2.23 & \cellcolor{Green!30}+13.49 \\
Neptune to& Chip-Seq & 4.99 & 4.66 & \cellcolor{Green!30}+7.16 \\
Jupiter& Nano-Seq & 0.19 & 0.39 & \cellcolor{Red!30}-52.06 \\
& Atac-Seq & 0.43 & 0.53 & \cellcolor{Yellow!30}-18.56 \\
\midrule
& RNA-Seq & 1.02 & 1.15 & \cellcolor{Green!30}-11.28 \\
Diana to& Chip-Seq & 1.17 & 1.54 & \cellcolor{Yellow!30}-23.87 \\
Neptune& Nano-Seq & 0.37 & 0.69 & \cellcolor{Red!30}-46.16 \\
& Atac-Seq & 0.78 & 1.08 & \cellcolor{Yellow!30}-27.75 \\
\midrule
& RNA-Seq & 2.66 & 2.23 & \cellcolor{Yellow!30}+19.19 \\
Diana to& Chip-Seq & 5.51 & 4.66 & \cellcolor{Yellow!30}+18.31 \\
Jupiter& Nano-Seq & 0.28 & 0.39 & \cellcolor{Red!30}-28.10 \\
& Atac-Seq & 0.52 & 0.53 & \cellcolor{Green!30}-0.11 \\
\midrule
& RNA-Seq & 0.96 & 1.15 & \cellcolor{Green!30}-16.47 \\
Minerva to& Chip-Seq & 1.51 & 1.54 & \cellcolor{Green!30}-1.53 \\
Neptune& Nano-Seq & 0.38 & 0.69 & \cellcolor{Red!30}-44.66 \\
& Atac-Seq & 0.96 & 1.08 & \cellcolor{Green!30}-10.79 \\
\midrule
& RNA-Seq & 2.09 & 2.23 & \cellcolor{Green!30}-6.16 \\
Minerva to& Chip-Seq & 5.95 & 4.66 & \cellcolor{Yellow!30}+27.79 \\
Jupiter& Nano-Seq & 0.21 & 0.39 & \cellcolor{Red!30}-46.85 \\
& Atac-Seq & 0.49 & 0.53 & \cellcolor{Green!30}-6.91 \\
\bottomrule
\end{tabular}
\end{table}

Augur's predicted energy consumption has a median error of 18.2\% (p75 of 27.9\% and p95 of 46.7\%). 
We noticed that Nano-Seq continued to exhibit the largest error, extrapolated by RAPL capturing all node energy consumption. 
Without Nano-Seq, our predictions would have a median error of 15.0\% (p75 of 18.5\% and p95 of 27.8\%) from RAPL readings.

\subsection{Prediction of Task Runtime}
\label{subsec:predict-task-runtime}

Because energy consumption is a derivative of execution duration, predicting task runtimes is a critical prerequisite for our model. 
We compared our predictions with Lotaru~\cite{bader2024lotaru}, a state-of-the-art profiling-based method to predict workflow task runtime. 

The results are shown in Table~\ref{table:augur-time-vs-lotaru}. 
Here, we sum the runtime of each workflow task into the overall workflow time, and compare this with the sum of each workflow task's predicted runtime. We show the prediction error, which is the percentage error between these two values. We also report the task prediction error, which is the mean error comparing each individual task's predicted runtime with its actual runtime. 

\begin{table*}[htb]
\caption{Comparison between using Augur and Lotaru to predict workflow task runtimes.}
\label{table:augur-time-vs-lotaru}
\centering
\begin{tabular}{ll|c|ccc|ccc}
\toprule
Profiling& & Actual & \multicolumn{3}{|c}{Augur} & \multicolumn{3}{|c}{Lotaru} \\
Node to& & Workflow & Predicted & Prediction & Task Prediction & Predicted & Prediction & Task Prediction \\
Target & Workflow & Time (s) & Time (s) & Error (\%) & Error (\%) & Time (s) & Error (\%) & Error (\%) \\
\midrule
& RNA-Seq & 52,744 & 74,153 & \cellcolor{green!30}+40.6 & +17.13 & 110,141 & +108.83 & +72.85 \\
Jupiter to&Chip-Seq & 64,044 & 63,532 & \cellcolor{green!30}-0.8 & +9.53 & 88,884 & +46.91 & +50.18 \\
Neptune&Nano-Seq & 25,533 & 19,629 & -23.12 & -22.64 & 28,970 & \cellcolor{green!30}+13.47 & +26.92 \\
& Atac-Seq & 47,816 & 62,109 & \cellcolor{green!30}+29.89 & +13.62 & 93,756 & +96.08 & +67.88 \\
\midrule
& RNA-Seq & 213,862 & 218,462 & \cellcolor{green!30}+2.15 & -11.76 & 142,909 & -30.01 & -32.54 \\
Neptune to&Chip-Seq & 339,535 & 324,374 & \cellcolor{green!30}-4.47 & -12.03 & 208,391 & -35.51 & -40.24 \\
Jupiter&Nano-Seq & 12,882 & 9,381 & \cellcolor{green!30}-27.17 & -12.09 & 6,205 & -51.83 & -33.4 \\
& Atac-Seq & 23,447 & 23,439 & \cellcolor{green!30}-0.02 & -0.15 & 17,827 & -23.96 & -27.12 \\
\midrule
& RNA-Seq & 54,718 & 49,706 & \cellcolor{green!30}-9.15 & -19.96 & 30,474 & -44.3 & -46.85 \\
Diana to&Chip-Seq & 64,044 & 51,203 & \cellcolor{green!30}-20.05 & -4.71 & 28,740 & -52.5 & -39.49 \\
Neptune&Nano-Seq & 25,533 & 16,850 & \cellcolor{green!30}-34.0 & -28.83 & 10,394 & -59.29 & -51.04 \\
& Atac-Seq & 47,816 & 32,965 & \cellcolor{green!30}-31.06 & -30.42 & 23,836 & -50.15 & -51.06 \\
\midrule
& RNA-Seq & 213,862 & 232,877 & \cellcolor{green!30}+8.89 & +20.28 & 53,622 & -73.74 & -68.71 \\
Diana to&Chip-Seq & 339,535 & 405,460 & \cellcolor{green!30}+19.42 & +34.98 & 93,567 & -71.05 & -65.96 \\
Jupiter&Nano-Seq & 12,882 & 14,137 & \cellcolor{green!30}+9.75 & +16.5 & 3,579 & -72.21 & -55.28 \\
& Atac-Seq & 23,447 & 26,084 & \cellcolor{green!30}+11.26 & +20.49 & 7,768 & -66.86 & -64.0 \\
\midrule
& RNA-Seq & 54,718 & 53,301 & \cellcolor{green!30}-2.58 & -20.39 & 46,607 & -14.82 & -29.36 \\
Minerva to&Chip-Seq & 64,044 & 65,299 & \cellcolor{green!30}+1.96 & -4.33 & 52,869 & -12.62 & -16.81 \\
Neptune&Nano-Seq & 25,533 & 17,999 & \cellcolor{green!30}-29.5 & -0.66 & 15,629 & -38.79 & -14.35 \\
& Atac-Seq & 47,816 & 42,188 & \cellcolor{green!30}-11.77 & -11.61 & 38,227 & -20.05 & -22.58 \\
\midrule
& RNA-Seq & 213,862 & 168,441 & \cellcolor{green!30}-21.24 & -10.36 & 84,673 & -58.53 & -46.97 \\
Minerva to&Chip-Seq & 339,535 & 371,050 & \cellcolor{green!30}+9.28 & -3.91 & 183,327 & -43.27 & -48.37 \\
Jupiter&Nano-Seq & 12,882 & 10,599 & \cellcolor{green!30}-17.72 & +6.81 & 5,491 & -57.37 & -35.73 \\
& Atac-Seq & 23,447 & 23,877 & \cellcolor{green!30}+1.84 & +6.35 & 12,479 & -46.77 & -43.47 \\
\bottomrule
\end{tabular}
\end{table*}

We see that Augur consistently predicts the overall workflow runtime with greater accuracy than Lotaru, and generally with a lower average task prediction error. We believe this is because the translation step used by Augur better captures the relationship between machines with a different number of cores, which is not considered by Lotaru. 

\subsection{Comparison with Baseline Method}  
\label{subsec:vs-baseline}
With limited existing methods available for predicting the energy consumption consumed by scientific workflows, we selected the approach by Barbosa et al.~\cite{queirozperformance} as a baseline. Their methodology requires historical execution traces and aligned energy consumption data to provide workflow-level predictions of energy consumption.
By analyzing workflow execution parameters, they established a relationship between the total number of bytes read and written by a workflow, and the total time spent reading and writing by a workflow with the energy consumed. They use this to train machine learning models to predict energy consumption. They found their Gradient Boosting regression (GBR) model to yield the most consistent results.

We adopted the approach provided by Barbosa et al. from their publicly available codebase. We varied the data pre-processing stage to train two model instances: For Barbosa--A, we provide the bytes read/written from 3 executions each of 3 workflows on the targeted compute cluster as training data. This is the same data provided to Augur with the reflector module enabled. In Barbosa--B, in addition to those 9 executions, there are a further 10 workflow executions from the same compute cluster~\cite{thamm2026nf}%
, given that their method expects significant historical data. Both approaches use GBR owing to it being the best performing model in the study.

The results are presented in Table~\ref{table:pred-energy-vs-baseline}, where we compare predictions made using Augur with predictions made using both Barbosa instances, and highlight the best performing approach in green. 
We observe that both baseline instances tend to overpredict energy consumption, with Augur producing more accurate estimations. 
This demonstrates the potential for an approach driven by workflow-specific training data, rather than relying on coarse workflow-level parameters such as the total number of bytes read and written over workflow execution, which are unknown prior to execution.

\begin{table}[htb]
\caption{Comparison between using Augur and Baseline method to predict workflow energy consumption on Jupiter.}
\label{table:pred-energy-vs-baseline}
\centering
\begin{tabular}{l|cccc}
\toprule
& \multicolumn{4}{c}{Energy Consumption (kWh)} \\
Workflow & RAPL & Augur & Barbosa--A & Barbosa--B \\
\midrule
RNA-Seq & 2.23 & \cellcolor{green!30}1.68 & 5.53 & 4.61 \\
Chip-Seq & 4.66 & \cellcolor{green!30}3.42 & 1.94 & 1.93 \\
Nano-Seq & 0.39 & \cellcolor{green!30}0.15 & 1.02 & 2.38 \\ 
Atac-Seq & 0.53 & \cellcolor{green!30}0.53 & 0.72 & 0.72 \\
\bottomrule
\end{tabular}
\end{table}

\subsection{Overhead}
\label{subsec:profiling-overhead}
We measure the overhead of Augur, considering the data, time, and energy overhead. For each workflow, we used three profiling runs, each with input data of different magnitude. We recorded the time taken for each profiling run, and estimated the energy consumed with Ichnos. We report the overall overhead of three profiling runs, on three different nodes, targeting the Neptune and Jupiter clusters (Table~\ref{table:oh-workflow-profiling}). 

\begin{table}[htb]
\caption{Overhead of overall workflow profiling on each node targeting Neptune and Jupiter. OH = overhead.}
\label{table:oh-workflow-profiling}
\centering
\setlength{\tabcolsep}{2pt}%
\begin{tabular}{lc|cc|cc|cc}
\toprule
Profiling & & \multicolumn{2}{c|}{Data} & \multicolumn{2}{c|}{Time} & \multicolumn{2}{c}{Energy} \\
Node & Workflow & (GB) & OH (\%) & (mins) & OH (\%) & (kWh) & OH (\%) \\
\midrule
\multicolumn{8}{c}{to Neptune} \\
\midrule
\multirow{4}{*}{Jupiter} & Atac-Seq & 1.23 & 11.08 & 36.77 & 3.80 & 0.08 & \cellcolor{Green!30}6.65 \\
& Chip-Seq & 0.31 & 3.07 & 50.97 & 4.03 & 0.09 & \cellcolor{Green!30}5.64 \\
& Nano-Seq & 1.15 & 12.20 & 18.02 & 2.07 & 0.03 & \cellcolor{Green!30}4.70 \\
& RNA-Seq & 0.73 & 5.06 & 92.41 & 9.15 & 0.17 & \cellcolor{Yellow!30}14.08 \\
\midrule
\multirow{4}{*}{Diana} & Atac-Seq & 1.23 & 11.08 & 39.59 & 4.09 & 0.13 & \cellcolor{Yellow!30}11.42 \\
& Chip-Seq & 0.31 & 3.07 & 50.91 & 4.03 & 0.15 & \cellcolor{Green!30}9.45 \\
& Nano-Seq & 1.15 & 12.20 & 24.60 & 2.83 & 0.07 & \cellcolor{Green!30}10.06 \\
& RNA-Seq & 0.73 & 5.06 & 104.05 & 10.30 & 0.30 & \cellcolor{Red!30}25.36 \\
\midrule
\multirow{4}{*}{Minerva} & Atac-Seq & 1.23 & 11.08 & 89.53 & 9.26 & 0.12 & \cellcolor{Yellow!30}10.49 \\
& Chip-Seq & 0.31 & 3.07 & 77.14 & 6.10 & 0.10 & \cellcolor{Green!30}6.39 \\
& Nano-Seq & 1.15 & 12.20 & 37.90 & 4.36 & 0.03 & \cellcolor{Green!30}4.54 \\
& RNA-Seq & 0.73 & 5.06 & 170.61 & 16.89 & 0.14 & \cellcolor{Yellow!30}11.41 \\
\bottomrule
\multicolumn{8}{c}{to Jupiter} \\
\midrule
\multirow{4}{*}{Neptune} & Atac-Seq & 1.23 & 11.08 & 170.82 & 38.06 & 0.23 & \cellcolor{Red!30}39.77 \\
& Chip-Seq & 0.31 & 3.07 & 157.55 & 6.51 & 0.21 & \cellcolor{Green!30}3.79 \\
& Nano-Seq & 1.15 & 12.20 & 43.60 & 12.06 & 0.06 & \cellcolor{Red!30}15.06 \\
& RNA-Seq & 0.73 & 5.06 & 276.09 & 27.44 & 0.35 & \cellcolor{Yellow!30}13.40 \\
\midrule 
\multirow{4}{*}{Diana} & Atac-Seq & 1.23 & 11.08 & 39.59 & 8.82 & 0.13 & \cellcolor{Red!30}22.34 \\
& Chip-Seq & 0.31 & 3.07 & 50.91 & 2.10 & 0.15 & \cellcolor{Green!30}2.79 \\
& Nano-Seq & 1.15 & 12.20 & 24.60 & 6.81 & 0.07 & \cellcolor{Red!30}19.29 \\
& RNA-Seq & 0.73 & 5.06 & 104.05 & 10.34 & 0.30 & \cellcolor{Yellow!30}11.50 \\
\midrule 
\multirow{4}{*}{Minerva} & Atac-Seq & 1.23 & 11.08 & 89.53 & 19.95 & 0.12 & \cellcolor{Red!30}20.52 \\
& Chip-Seq & 0.31 & 3.07 & 77.14 & 3.19 & 0.10 & \cellcolor{Green!30}1.89 \\
& Nano-Seq & 1.15 & 12.20 & 37.90 & 10.49 & 0.03 & \cellcolor{Green!30}8.70 \\
& RNA-Seq & 0.73 & 5.06 & 170.61 & 16.95 & 0.14 & \cellcolor{Green!30}5.18 \\
\bottomrule
\end{tabular}
\end{table}

When predicting for the Neptune cluster, we see that using the Diana node, a powerful AWS metal instance, led to a worst-case time overhead of 10.3\% and energy overhead of 25.4\%. The best node for profiling was Jupiter, with a worst-case time overhead of 9.2\% and energy overhead of 14.1\%. 
For the Jupiter cluster, we saw that Neptune had the worst-case time overhead of 38.1\% and energy overhead of 39.8\%. 

On average, in Table~\ref{table:oh-workflow-profiling}, we had a median energy overhead of $10.3\% \pm 8.5\%$ (p75 14.3\% and p95 24.9\%). 
On the best performing profiling node for the execution on Neptune, which was from Jupiter, the energy overhead was as low as 4.7\% in the best-case, with the median energy overhead just 6.1\%.
Generally, we found that the overhead of workflow profiling was greater for workflows that had a smaller overall footprint. For example, the Nano-Seq workflow consumed 0.39$kWh$ on Jupiter, and the profiling runs consumed 9--19\% of the overall energy consumption.

\subsection{Discussion}
\label{subsec:discussion}
We discuss some of the challenges faced by our proposed approach. 

\paragraph{Unseen Workflow Tasks}
Given that dynamic workflow systems like Nextflow can branch based on input data, it is possible that we encounter an unseen task in a full sized workflow execution that is not present in profiling executions, as exemplified by the Nano-Seq workflow in our evaluation. 
In this scenario, we are reliant on access to limited historical data from the targeted infrastructure to make a prediction. 
If this is not available, there is no information to go from. Here, we would have to rely on some generic task runtime and CPU utilization, and therefore, energy prediction, which will likely introduce a significant estimation error. 

\paragraph{Profiling Overhead}
The intended application of an energy prediction method for tasks would be to execute a scientific workflow in an energy- or carbon-aware manner.
If energy-efficient scheduling algorithms that rely on prior knowledge of tasks are able to reduce energy consumption by 40 or 50\%~\cite{https://doi.org/10.1155/2022/1637614, reddy2023multi}, then a prediction method with an overhead of around 10\% would allow for significant savings to be made. 
However, it might not always be possible to make significant energy savings and the overhead might outweigh savings in some cases.
Here, the choice of profiling node would be important, ensuring that the node has enough memory to run workflow profiling, but that the time overhead and energy overhead are minimized.

However, there are opportunities available when minimizing the overhead for carbon-aware execution. For example, given that the profiling is done on a single node and does not require the user to collect energy consumption data, a scientist could provision a cloud node for this in a region where the grid runs on low-carbon energy sources, reducing the carbon footprint of profiling runs, even if the same amount of energy is consumed. This would minimize the overhead while predicting runtimes and energy consumptions for carbon-aware scheduling, overall reducing the workflow footprint.

\section{Related Work}
In this section, we first discuss works focused on energy- and carbon-aware workflow scheduling, where predicted task runtime and energy consumption would be required (in Section~\ref{sec:rw-wf-scheduling}). 
Then, we discuss runtime prediction methods for scientific workflows (in Section~\ref{sec:rw-predict-runtime}), followed by work focused on estimating the energy consumption of scientific workflow executions based on metrics or from monitoring their execution (in Section~\ref{sec:rw-estimate-energy}).
Finally, we discuss previous work on energy consumption prediction (in Section~\ref{subsec:previous-energy-prediction}). 

\subsection{Energy- and Carbon-Aware Workflow Scheduling}  
\label{sec:rw-wf-scheduling}
In recent years, there has been an increase in work focused on reducing the energy consumption~\cite{reddy2023multi, durillo2014multi, xu2015enreal}, and carbon footprint of scientific workflows~\cite{vivas2024trends, bostandoost2025quantifyingcarbonreductiondag, 9298863, 10.1145/3754598.3754673, west2026systematic}. 

Reddy et al.~\cite{reddy2023multi} reduce the energy consumption of workflows by clustering tasks according to their runtimes. 
Durillo et al.~\cite{durillo2014multi} uses neural networks to predict execution time and energy consumption, assuming significant historical data is available to train  models to be applied for energy-efficient workflow scheduling. 
Xu et al.~\cite{xu2015enreal} proposed an energy-aware resource allocation method for scientific workflows in the cloud, where they assumed that task durations were known in advance. 
Bostandoost et al.~\cite{bostandoost2025quantifyingcarbonreductiondag} explored how workflows can be scheduled to reduce carbon emissions, assuming that task durations are known in advance on machines with set, known power draws. 
Wen et al.~\cite{9298863} created an algorithm to increase green energy usage when scheduling industrial workflows. Their evaluation assumed that the rate that VM energy consumption is constant and known, and that the execution time on each device is known. 
Schweisgut et al.~\cite{10.1145/3754598.3754673} proposed a scheduling algorithm to reduce the carbon emissions from workflow executions by aligning their execution with low carbon energy. Each task has a computation and communication time, and both were known in advance. Each processor had a known idle and active power consumption unit, consumed over each time step. 
West et al.~\cite{west2026systematic} explored how the carbon footprint of scientific workflows can be reduced. They simulated workflow execution while applying carbon-aware techniques, and assumed that the workflow task runtimes and energy consumptions were known.

Given that these works typically assumed that they had a-priori information on workflow task runtimes and energy consumption when making optimizations, they would benefit from pre-execution predictions of energy consumption to work in practice. 

\subsection{Prediction of Runtime for Scientific Workflows}
\label{sec:rw-predict-runtime}
Several methods have been developed to predict the runtime of workflow tasks, which can be separated according to when they train their prediction models.

\paragraph{Offline methods} These methods create models using historical data, before workflows are executed.
Several methods train neural networks to predict the runtime of applications and workflows~\cite{nadeem2017modeling, malik2013execution, huang2024cloudprophet}. 
These approaches assume that historical data is available to train and validate their models. However, this is not always the case when scientists execute new workflows, on different infrastructure. Training large models can also require hundreds of compute hours, leading to significant energy consumption. 
Given that our approach is intended to work with limited historical data to make predictions also for, so far, unseen workflows, or workflows running with new input data, we do not consider these methods applicable. 

\paragraph{Online methods} These methods train and retrain their models while applications are executed. 
Da Silva et al.~\cite{10.1145/2534248.2534254, da2015online} build regression trees used to predict workflow task resource usage and execution times. 
Hilman et al.~\cite{hilman2018task} use short term memory networks to predict workflow task execution times in the cloud.
Both approaches continually adjust their models at runtime, as tasks are executed. 
Given that both approaches require initial access to historical workflow data, we do not consider them applicable. 

\paragraph{Profiling-based methods} These methods predict workflow task execution times by profiling on input samples and limited resources. Lotaru~\cite{bader2024lotaru} performed profiling on a user's local machine and targeted infrastructure, running workflows with down-sampled data to collect task metrics used to identify task behavior, and predict the runtime with a Bayesian linear regression model for targeted machines. 
We base our approach on Lotaru, and implement it as a baseline to compare against in our evaluation. 

\subsection{Estimating the Energy Consumption of Scientific Workflow Execution}
\label{sec:rw-estimate-energy}
Some methods have been developed to estimate the energy consumption for a scientific workflow execution, either based on \textit{collected metrics} or from \textit{measurements during execution}. 
We distinguish energy consumption estimation from prediction, as this estimation occurs \textit{after} execution. 

\paragraph{Estimation from Collected Metrics}
Ichnos~\cite{west2025ichnoscarbonfootprintestimator} estimates the energy consumed, for scientific workflow tasks based on the runtime, the CPU utilization, and a power model generated from stress-testing the CPU and memory to improve accuracy. 
We adopt the same approach as Ichnos, using the predicted task runtime, CPU utilization, and memory to estimate the energy consumption using power models acquired from the infrastructure profiling phase.

Some methodologies estimate the energy consumption of a workload in order to estimate carbon emissions. Green Algorithms (GA)\cite{lannelongue2021Green} uses the runtime, CPU utilization, memory, and the CPU's reported thermal design power, while Cloud Carbon Footprint\textsuperscript{\ref{mark:ccf}} uses the runtime, CPU utilization, memory, and a linear power model formed from a CPU's minimum and maximum power consumption. 
The nf-co2footprint plugin\footnote{\url{https://github.com/nextflow-io/nf-co2footprint}} implements the GA method for scientific workflows as a Nextflow plugin. 

\paragraph{Estimation from Measurements during Execution}
Other approaches can estimate energy consumption by monitoring their execution. 
Warade et al.~\cite{warade2022measuring} monitor the energy consumed over workflow execution, showing the potential for energy-aware scheduling to be applied. 
Thamm et al.~\cite{thamm2025strategiesmeasureenergyconsumption} monitor energy consumed by reading Intel RAPL energy counters, and they discuss an approach for task-based measurement for tasks that run in isolation, for tasks that run concurrently they note that heuristics could be used to estimate total energy consumption caused by individual tasks. 

While directly monitoring workflow execution would produce the most accurate values~\cite{khan2018rapl}, we do not have this monitoring data prior to workflow execution, and typically Nextflow applications involve significant concurrency across utilized resources. Therefore, we predict energy consumption based on available metrics. 

\subsection{Energy Consumption Prediction}
\label{subsec:previous-energy-prediction}
Several works have considered how the energy consumed by compute workloads running on HPC infrastructure could be predicted~\cite{saad2025carbonawarecontainerorchestrationpredicting,8048921,10603495,9139801}. 

Some approaches rely on having access to significant historical job submission data, detailing characteristics like power consumption, CPU utilization, the number of nodes utilized, and the user that made the submission~\cite{10603495, 9139801}. 
Others combine these characteristics with energy consumption data, to predict energy consumption for simulated benchmarks~\cite{saad2025carbonawarecontainerorchestrationpredicting} or for  MPI applications~\cite{8048921}, where the workloads are assumed to exhibit known and repeatable behavior. 

Other works have instead focused on the cost of executing workflow tasks on cloud infrastructure~\cite{rosa2018cost, rosa2021computational}. They gather training data from repeatedly running workflows, or simulating their execution to make predictions using a multiple linear regression model. They make predictions for tasks that run in isolation on a VM instance and do not consider tasks that run concurrently. 

There has been little work that focuses on predicting the energy consumed by scientific workflows. 
Barbosa et al.~\cite{queirozperformance} use a significant amount of historical workflow traces and corresponding energy consumption data to train models to predict the energy consumption of workflow executions. This prediction is only made at a workflow-level. 
Da Silva et al.~\cite{da2020characterizing} estimate the energy consumption of I/O-intensive scientific workflow tasks by training a power model that considers both CPU and I/O energy consumption. They repeatedly executed tasks on targeted infrastructure, both in isolation and in parallel, to collect energy consumption data alongside task execution data. Such required significant training time, which would incur significant energy overhead. Furthermore, it required that users had access to directly monitor energy consumption, which is not always possible in cloud environments, or restricted clusters. 

Crucially, both approaches rely on having significant historical data, with aligning energy consumption data. The scientific workflows we aim to predict the energy consumption for may be new, or run on different infrastructure.
In contrast, Augur predicts workflow task energy consumption based on workflow profiling runs executed on a single node, and does not require energy consumption data to be gathered. 

\section{Conclusion} 
In this paper, we presented Augur, a novel approach for predicting the energy consumption of scientific workflow tasks on heterogeneous clusters based on single-node profiling. Our evaluation against established baselines demonstrated that the overall scientific workflow energy consumption can be predicted with a median error of $16.3\% \pm 15.3\%$ when compared to Ichnos, and $18.2 \pm 14.7\%$ compared to RAPL. We also compared task runtime predictions, showing that Augur consistently outperformed Lotaru in predictive accuracy for utilized infrastructure. Furthermore, we compared the predicted workflow energy consumption with Barbosa's approach, showing that Augur consistently made predictions with greater accuracy. Notably, we critically discussed the challenges faced by our approach, emphasizing the necessary balance between profiling overheads and the achievable reductions in energy consumption and carbon emissions.

Future work will explore the applications of Augur, using the predicted runtime and energy consumption of scientific workflow tasks to schedule their execution in an energy-efficient, carbon-aware manner with a resource manager. Integrating these predictions with real-time carbon intensity forecasts would represent a transformative step towards carbon-aware scheduling algorithms tailored to scientific workflows. 

\section*{Acknowledgment}
This work was supported by the Engineering and Physical Sciences Research Council under grant number UKRI154 (``Casper: Carbon-Aware Scalable Processing in Elastic Clusters'') and the German Research Council (DFG) as part of the CRC 1404 ( ``FONDA: Foundations of Workflows for Large-Scale Scientific Data Analysis''). 
We thank AWS and Google Cloud for providing research cloud credits to support this work.

\balance
\bibliographystyle{IEEEtran}
\bibliography{references}

\begin{thebibliography}{10}
\providecommand{\url}[1]{#1}
\csname url@samestyle\endcsname
\providecommand{\newblock}{\relax}
\providecommand{\bibinfo}[2]{#2}
\providecommand{\BIBentrySTDinterwordspacing}{\spaceskip=0pt\relax}
\providecommand{\BIBentryALTinterwordstretchfactor}{4}
\providecommand{\BIBentryALTinterwordspacing}{\spaceskip=\fontdimen2\font plus
\BIBentryALTinterwordstretchfactor\fontdimen3\font minus \fontdimen4\font\relax}
\providecommand{\BIBforeignlanguage}[2]{{%
\expandafter\ifx\csname l@#1\endcsname\relax
\typeout{** WARNING: IEEEtran.bst: No hyphenation pattern has been}%
\typeout{** loaded for the language `#1'. Using the pattern for}%
\typeout{** the default language instead.}%
\else
\language=\csname l@#1\endcsname
\fi
#2}}
\providecommand{\BIBdecl}{\relax}
\BIBdecl

\bibitem{muirRealCostSequencing2016b}
P.~Muir, S.~Li, S.~Lou, D.~Wang, D.~J. Spakowicz, L.~Salichos, J.~Zhang, G.~M. Weinstock, F.~Isaacs, J.~Rozowsky, and M.~Gerstein, ``The real cost of sequencing: Scaling computation to keep pace with data generation,'' \emph{Genome Biology}, vol.~17, no.~1, 2016.

\bibitem{fellowsyatesReproduciblePortableEfficient2021}
J.~A. Fellows~Yates \emph{et~al.}, ``Reproducible, portable, and efficient ancient genome reconstruction with nf-core/eager,'' \emph{PeerJ}, vol.~9, 2021.

\bibitem{schaarschmidtWorkflowEngineeringMaterials2021}
J.~Schaarschmidt \emph{et~al.}, ``Workflow {{Engineering}} in {{Materials Design}} within the {{BATTERY}} 2030 + {{Project}},'' \emph{Advanced Energy Materials}, vol.~12, 2021.

\bibitem{berrimanMontageGridEnabled2004}
B.~Berriman \emph{et~al.}, ``Montage: {{A}} grid enabled engine for delivering custom science-grade mosaics on demand,'' in \emph{Optimizing Scientific Return for Astronomy through Information Technologies}, vol. 5493, 2004.

\bibitem{ditommasoNextflowEnablesReproducible2017a}
P.~Di~Tommaso, M.~Chatzou, E.~W. Floden, P.~P. Barja, E.~Palumbo, and C.~Notredame, ``Nextflow enables reproducible computational workflows,'' \emph{Nature biotechnology}, vol.~35, no.~4, 2017.

\bibitem{freitagRealClimateTransformative2021b}
C.~Freitag, M.~Berners-Lee, K.~Widdicks, B.~Knowles, G.~S. Blair, and A.~Friday, ``The real climate and transformative impact of {{ICT}}: {A} critique of estimates, trends, and regulations,'' \emph{Patterns}, vol.~2, no.~9, 2021.

\bibitem{vivas2024trends}
A.~Vivas, A.~Tchernykh, and H.~Castro, ``Trends, approaches, and gaps in scientific workflow scheduling: A systematic review,'' \emph{IEEE Access}, vol.~12, 2024.

\bibitem{da2024workflows}
R.~F. Da~Silva, D.~Bard, K.~Chard, S.~De~Witt, I.~T. Foster, T.~Gibbs, C.~Goble, W.~Godoy, J.~Gustafsson, U.-U. Haus \emph{et~al.}, ``Workflows community summit 2024: Future trends and challenges in scientific workflows,'' \emph{arXiv}, 2024.

\bibitem{reddy2023multi}
P.~V. Reddy and K.~G. Reddy, ``A multi-objective based scheduling framework for effective resource utilization in cloud computing,'' \emph{IEEE Access}, vol.~11, 2023.

\bibitem{durillo2014multi}
J.~J. Durillo, V.~Nae, and R.~Prodan, ``Multi-objective energy-efficient workflow scheduling using list-based heuristics,'' \emph{Future Generation Computer Systems}, vol.~36, 2014.

\bibitem{bostandoost2025quantifyingcarbonreductiondag}
R.~Bostandoost, A.~Lechowicz, W.~A. Hanafy, P.~Shenoy, and M.~Hajiesmaili, ``Quantifying the carbon reduction of {DAG} workloads: A job shop scheduling perspective,'' \emph{arXiv}, 2025.

\bibitem{west2026systematic}
K.~West, Y.~Moawad, F.~Lehmann, V.~Bountris, U.~Leser, Y.~Elkhatib, and L.~Thamsen, ``A systematic evaluation of the potential of carbon-aware execution for scientific workflows,'' \emph{Future Generation Computer Systems}, 2026.

\bibitem{10.1145/3754598.3754673}
D.~Schweisgut, A.~Benoit, Y.~Robert, and H.~Meyerhenke, ``Carbon-aware workflow scheduling with fixed mapping and deadline constraint,'' in \emph{Proceedings of the 54th International Conference on Parallel Processing (ICPP)}.\hskip 1em plus 0.5em minus 0.4em\relax ACM, 2025.

\bibitem{9298863}
Z.~Wen, S.~Garg, G.~S. Aujla, K.~Alwasel, D.~Puthal, S.~Dustdar, A.~Y. Zomaya, and R.~Ranjan, ``Running industrial workflow applications in a software-defined multicloud environment using green energy aware scheduling algorithm,'' \emph{IEEE Transactions on Industrial Informatics}, vol.~17, no.~8, 2021.

\bibitem{durillo2015pareto}
J.~J. Durillo, R.~Prodan, and J.~G. Barbosa, ``Pareto tradeoff scheduling of workflows on federated commercial clouds,'' \emph{Simulation Modelling Practice and Theory}, vol.~58, 2015.

\bibitem{bader2024lotaru}
J.~Bader, F.~Lehmann, L.~Thamsen, U.~Leser, and O.~Kao, ``Lotaru: Locally predicting workflow task runtimes for resource management on heterogeneous infrastructures,'' \emph{Future Generation Computer Systems}, vol. 150, 2024.

\bibitem{da2015online}
R.~F. Da~Silva, G.~Juve, M.~Rynge, E.~Deelman, and M.~Livny, ``Online task resource consumption prediction for scientific workflows,'' \emph{Parallel Processing Letters}, vol.~25, no.~3, 2015.

\bibitem{rosa2018cost}
M.~J. Rosa, A.~P. Ara{\"u}jo, and F.~L. Mendes, ``Cost and time prediction for efficient execution of bioinformatics workflows in federated cloud,'' in \emph{International Conference on Bioinformatics and Biomedicine (BIBM)}.\hskip 1em plus 0.5em minus 0.4em\relax IEEE, 2018.

\bibitem{rosa2021computational}
M.~J. Rosa, C.~G. Ralha, M.~Holanda, and A.~P. Araujo, ``Computational resource and cost prediction service for scientific workflows in federated clouds,'' \emph{Future Generation Computer Systems}, vol. 125, 2021.

\bibitem{da2020characterizing}
R.~F. da~Silva, H.~Casanova, A.-C. Orgerie, R.~Tanaka, E.~Deelman, and F.~Suter, ``Characterizing, modeling, and accurately simulating power and energy consumption of i/o-intensive scientific workflows,'' \emph{Journal of computational science}, vol.~44, 2020.

\bibitem{queirozperformance}
F.~Barbosa, E.~Damasceno, F.~Queiroz, J.~Brito, J.~de~Souza~Ara{\'u}jo, and M.~Amaris, ``Performance and energy consumption prediction of scientific workflows using machine learning,'' in \emph{Latin American High Performance Computing Conference}, 2026.

\bibitem{ewels2020nf}
P.~A. Ewels, A.~Peltzer, S.~Fillinger, H.~Patel, J.~Alneberg, A.~Wilm, M.~U. Garcia, P.~Di~Tommaso, and S.~Nahnsen, ``The nf-core framework for community-curated bioinformatics pipelines,'' \emph{Nature biotechnology}, vol.~38, no.~3, 2020.

\bibitem{west2025ichnoscarbonfootprintestimator}
K.~West, M.~Reid, Y.~Elkhatib, and L.~Thamsen, ``Ichnos: A carbon footprint estimator for scientific workflows,'' in \emph{LOCO Workshop}, 2025.

\bibitem{khan2018rapl}
K.~N. Khan, M.~Hirki, T.~Niemi, J.~K. Nurminen, and Z.~Ou, ``{RAPL} in action: Experiences in using {RAPL} for power measurements,'' \emph{ACM Transactions on Modeling and Performance Evaluation of Computing Systems (TOMPECS)}, vol.~3, no.~2, 2018.

\bibitem{thamm2025strategiesmeasureenergyconsumption}
P.~Thamm and U.~Leser, ``Strategies to measure energy consumption using {RAPL} during workflow execution on commodity clusters,'' \emph{arXiv}, 2025.

\bibitem{thamm2026nf}
P.~Thamm, S.~Mohammadi, K.~West, K.~Reinert, L.~Thamsen, and U.~Leser, ``Nf-peak: Process-based energy attribution for nextflow workflows on kubernetes clusters,'' \emph{arXiv preprint arXiv:2605.22393}, 2026.

\bibitem{https://doi.org/10.1155/2022/1637614}
N.~Garg, Neeraj, M.~Raj, I.~Gupta, V.~Kumar, and G.~R. Sinha, ``Energy-efficient scientific workflow scheduling algorithm in cloud environment,'' \emph{Wireless Communications and Mobile Computing}, vol. 2022, no.~1, 2022.

\bibitem{xu2015enreal}
X.~Xu, W.~Dou, X.~Zhang, and J.~Chen, ``{EnReal}: An energy-aware resource allocation method for scientific workflow executions in cloud environment,'' \emph{IEEE Transactions on Cloud Computing}, vol.~4, no.~2, 2015.

\bibitem{nadeem2017modeling}
F.~Nadeem, D.~Alghazzawi, A.~Mashat, K.~Fakeeh, A.~Almalaise, and H.~Hagras, ``Modeling and predicting execution time of scientific workflows in the grid using radial basis function neural network,'' \emph{Cluster Computing}, vol.~20, no.~3, 2017.

\bibitem{malik2013execution}
M.~J. Malik, T.~Fahringer, and R.~Prodan, ``Execution time prediction for grid infrastructures based on runtime provenance data,'' in \emph{Proceedings of the 8th Workshop on Workflows in Support of Large-Scale Science (WORKS)}, 2013.

\bibitem{huang2024cloudprophet}
D.~Huang, L.~Costero, A.~Pahlevan, M.~Zapater, and D.~Atienza, ``Cloudprophet: a machine learning-based performance prediction for public clouds,'' \emph{IEEE Transactions on Sustainable Computing}, vol.~9, no.~4, 2024.

\bibitem{10.1145/2534248.2534254}
R.~F. da~Silva, G.~Juve, E.~Deelman, T.~Glatard, F.~Desprez, D.~Thain, B.~Tovar, and M.~Livny, ``Toward fine-grained online task characteristics estimation in scientific workflows,'' in \emph{Proceedings of the 8th Workshop on Workflows in Support of Large-Scale Science (WORKS)}, 2013.

\bibitem{hilman2018task}
M.~H. Hilman, M.~A. Rodriguez, and R.~Buyya, ``Task runtime prediction in scientific workflows using an online incremental learning approach,'' in \emph{IEEE/ACM 11th International Conference on Utility and Cloud Computing (UCC)}, 2018.

\bibitem{lannelongue2021Green}
L.~Lannelongue, J.~Grealey, and M.~Inouye, ``Green algorithms: Quantifying the carbon footprint of computation,'' \emph{Advanced science}, vol.~8, no.~12, 2021.

\bibitem{warade2022measuring}
M.~Warade, J.-G. Schneider, and K.~Lee, ``Measuring the energy and performance of scientific workflows on low-power clusters,'' \emph{Electronics}, vol.~11, no.~11, 2022.

\bibitem{saad2025carbonawarecontainerorchestrationpredicting}
Z.~Saad, J.~Yang, H.~Leung, and S.~Drew, ``Towards carbon-aware container orchestration: Predicting workload energy consumption with federated learning,'' in \emph{IEEE Smart World Congress (SWC)}, 2025.

\bibitem{8048921}
F.~C. Heinrich, T.~Cornebize, A.~Degomme, A.~Legrand, A.~Carpen-Amarie, S.~Hunold, A.-C. Orgerie, and M.~Quinson, ``Predicting the energy-consumption of {MPI} applications at scale using only a single node,'' in \emph{IEEE International Conference on Cluster Computing (CLUSTER)}, 2017.

\bibitem{10603495}
Y.~Lou, J.~Wang, S.~Feng, and X.~Yu, ``An efficient energy consumption prediction framework for high performance computing cluster jobs,'' in \emph{5th International Conference on Computer Engineering and Application (ICCEA)}, 2024.

\bibitem{9139801}
T.~Patel, A.~Wagenhäuser, C.~Eibel, T.~Hönig, T.~Zeiser, and D.~Tiwari, ``What does power consumption behavior of {HPC} jobs reveal? {D}emystifying, quantifying, and predicting power consumption characteristics,'' in \emph{IEEE International Parallel and Distributed Processing Symposium (IPDPS)}, 2020.

\end{thebibliography}

\end{document}